\documentclass[journal,twoside,web]{ieeecolor}
\usepackage{tmi}
\usepackage{cite}
\usepackage{amsmath,amssymb,amsfonts}
\usepackage{algorithmic}
\usepackage{graphicx}
\usepackage{textcomp}
\usepackage{booktabs}
\usepackage{url} 
\usepackage{bm,xspace}
\usepackage{amsmath, esvect}

\newcommand{\xmath}[1] {\ensuremath{#1}\xspace}
\newcommand{\blmath}[1] {\xmath{\bm{#1}}}

\newcommand{\A} {\blmath{A}}
\newcommand{\W} {\blmath{W}}
\newcommand{\x} {\blmath{x}}
\newcommand{\y} {\blmath{y}}
\newcommand{\rr} {\blmath{r}}
\newcommand{\bmtheta} {\blmath{\theta}}
\newcommand{\bmbeta} {\blmath{\beta}}

\DeclareMathOperator*{\argmin}{arg\,min}

\long\def\red#1{\bgroup\color{red}#1\egroup}
\newcommand{\defequ}{\triangleq}
\def\BibTeX{{\rm B\kern-.05em{\sc i\kern-.025em b}\kern-.08em
    T\kern-.1667em\lower.7ex\hbox{E}\kern-.125emX}}
\begin{document}
\bstctlcite{IEEEexample:BSTcontrol}
\title{Bilevel Optimized Implicit Neural Representation for Scan-Specific Accelerated 
MRI Reconstruction}
\author{Hongze Yu, Jeffrey A. Fessler, and Yun Jiang
\thanks{This work was supported by NIH grants R37CA263583, R01CA284172, 
and Siemens Healthineers. }
\thanks{This work involved human subjects or animals in its research.
The study was HIPAA-compliant and was approved by the Institutional Review Board.
Informed written consent was obtained from each participant.}
\thanks{Hongze Yu and Jeffrey A. Fessler are with the Department of Electrical Engineering
and Computer Science, University of Michigan, Ann Arbor, MI 48109 USA 
(e-mail: hongze@umich.edu; fessler@umich.edu).}
\thanks{Yun Jiang is with the Department of Radiology and the Department of Biomedical Engineering,
University of Michigan, Ann Arbor, MI 48109 USA (e-mail: yunjiang@umich.edu).}}

\maketitle
\thispagestyle{empty}

\begin{abstract}
Deep Learning (DL) methods can reconstruct highly accelerated
magnetic resonance imaging (MRI) scans,
but they rely on application-specific large training datasets
and often generalize poorly to out-of-distribution data.
Self-supervised deep learning algorithms perform scan-specific reconstructions,
but still require complicated hyperparameter tuning
based on the acquisition and often offer limited acceleration.
This work develops a bilevel-optimized implicit neural representation (INR) approach
for scan-specific MRI reconstruction.
The method automatically optimizes the hyperparameters
for a given acquisition protocol,
enabling a tailored reconstruction without training data.
The proposed algorithm uses Gaussian process regression to optimize INR hyperparameters,
accommodating various acquisitions.
The INR includes a trainable positional encoder for high-dimensional feature embedding
and a small multilayer perceptron for decoding.
The bilevel optimization is computationally efficient, requiring only a few minutes per typical 2D Cartesian scan. On scanner hardware, the subsequent scan-specific reconstruction—using offline-optimized hyperparameters—is completed within seconds and achieves improved image quality compared to previous model-based and self-supervised learning methods.

\end{abstract}


\section{Introduction}
\label{sec:introduction}

\IEEEPARstart{M}{agnetic} resonance imaging (MRI)
is a diagnostic imaging technique that provides excellent soft-tissue contrast
without ionizing radiation.
However, its prolonged acquisition times can pose a challenge in clinical practice.
To address this drawback,
a common strategy is to develop reconstruction algorithms
that can recover artifact-free images from sub-Nyquist sampled data.
Parallel imaging techniques \cite{SENSE, GRAPPA} exploit redundancy among multiple receiver coils.
Classical methods, however, are constrained by the number geometry of the coils,
leading to potential residual artifacts and noise amplification at high acceleration rates.
Compressed sensing~\cite{sparse_mri,compressed_sensing_TV} and related 
model-based methods~\cite{LORAKS,PLORAKS,annihilating_filter_low_rank}
further mitigate aliasing artifacts by enforcing sparsity or low-rank constraints
and can be combined with parallel imaging
to achieve higher acceleration rates in dynamic and quantitative MRI%
~\cite{ktFOCUSS,GRASP,XDGRASP,LLR_MRF}.
However, their computational overhead limits their broader clinical adoption~\cite{ye_mri_review}.

Supervised deep learning-based methods can offer efficient reconstruction
%directly at the MRI scanner
once they are trained.
Typically, neural networks are trained using a large fully sampled dataset~\cite{fastMRI}.
Common approaches include training the neural network
to learn a direct mapping from undersampled images to fully sampled images
\cite{parallel_mlp, dl_mri_mapping},
or employing the network as a regularizer in model-based iterative reconstruction~\cite{modl, ista_net}.
However, these require curated fully sampled training datasets~\cite{knoll_deeplearning}
that are often unavailable for many MRI acquisitions, such as dynamic or diffusion imaging.
Moreover, these pre-trained models often struggle to generalize effectively
when applied to out-of-distribution data~\cite{OOD_mri}.

Various semi- and self-supervised deep learning-based methods have been proposed
to overcome the need for fully sampled data in supervised methods,
focusing on scan-specific reconstruction.
For instance,
Noise2Recon~\cite{noise2recon} enforces consistency
between model reconstructions of undersampled scan data and their noise-augmented counterparts,
while self-supervised learning via data undersampling (SSDU)~\cite{SSDU}
divides undersampled data into subsets to ensure data consistency and calculate training losses.
Other approaches \cite{DIP,DNLINV} rely on implicit regularization through specific network architectures.

Recent advances in novel-view synthesis and volume rendering~\cite{nerf} offer a new approach
to reconstructing undersampled MRI data.
INRs model target signals, such as MR images,
as continuous coordinate-based functions, inherently enforcing continuity. 
In novel-view synthesis, INR learns from multiple projection views
to reconstruct a continuous target scene.
Similarly, in MRI, INR can leverage the redundancy of coil sensitivity
to reconstruct alias-free images from undersampled MRI data. 
Previous works have achieved scan-specific INR-based reconstruction
by employing fully sampled images as priors~\cite{NeRP}
or adding explicit, hand-crafted regularizers such as total variation and low rank \cite{IMJENSE, CEST_INR, spatiotemporal_inr}.
However, INR-based methods, like other self-supervised reconstruction techniques,
are sensitive to the choice of hyperparameters.
These methods typically require dedicated hyperparameter tuning (e.g., grid or random searches) tailored to specific datasets,
considering factors such as anatomy, contrast, sampling pattern.
Poorly tuned hyperparameters can directly degrade reconstructed image quality.

In this work, we propose a \emph{bilevel-optimized INR}
for hyperparameter-optimized, scan-specific MRI reconstruction.
We introduce a self-regularized INR network that achieves high-quality reconstructions
without requiring external priors
by systematically designing the key components,
such as positional encoder, decoder multilayer perceptron (MLP) and loss function weighting.
To enable automatic hyperparameter tuning,
we formulate the reconstruction as a bilevel optimization problem~\cite{bilevel_image_recon},
where the lower level is a self-supervised INR network
and the upper level handles hyperparameter optimization,
solved using Bayesian optimization~\cite{bayesopt}. 
We divide the undersampled single-scan k-space data
into mutually exclusive training and validation sets to enable self-supervision.
For each candidate hyperparameter vector,
the INR is trained on the training set, and the validation loss is computed using unseen data.
We adopted Bayesian optimization for its derivative-free characteristic and computational efficiency,
making it suitable for this global optimization task.
The proposed bilevel-optimized INR can adapt hyperparameters to various acquisition protocols,
such as anatomy, sampling pattern, contrast.
We also demonstrate that these hyperparameters are transferable
between subjects or similar anatomies scanned using the same acquisition protocol.
Our proposed method improves image quality
compared with images reconstructed using state-of-the-art self-supervised deep-learning
and model-based methods.
With offline-optimized hyperparameters,
our approach can achieve scan-specific MRI reconstruction,
in under 6 seconds for a 2D Cartesian scan
with matrix size \(384\times 384\) on a single NVIDIA A40 GPU.
Non-Cartesian reconstruction is approximately 5$\times$ slower
due to non-uniform fast Fourier transform (NUFFT) computation.

We also validate hyperparameter transferability across same anatomies
using interventional MRI (iMRI) with residual learning.
Our INR method, hyperparameter-optimized on the same anatomy
and trained on the fully sampled first frame of an MR-guided biopsy scan,
reconstructs subsequent undersampled frames with minimal fine-tuning
(under 1 second per frame with a matrix size of \(128\times 128\)).
This residual learning strategy potentially enables real-time iMRI 
reconstruction with highly accelerated acquisition (\(6\times\)).

The main contributions of this study are: 
(1) we propose a scan-specific bilevel-optimized INR for reconstructing 
undersampled MR data, enabling automatic hyperparameter tuning for each scan;
(2) we systematically investigate design choices for key INR components, such as the positional encoder, MLP decoder, and loss function, and demonstrate that with proper tuning, 
the INR-based reconstruction can achieve high reconstruction quality without explicit regularization;
(3) we demonstrate that optimized hyperparameters are transferrable across subjects or similar anatomies
scanned with the same acquisition using bilevel-optimized INR, while also validating 
the necessity of tailored hyperparameter tuning for different acquisitions;
(4) with hyperparameters optimized offline, our method achieves tailored 
reconstructions at computation times under 6 seconds, indicating the potential  
toward clinical translation of self-supervised deep learning reconstructions;
(5) we validate the proposed algorithm across a wide range of MR applications, 
including various anatomies, image contrasts, field strengths, and sampling patterns.

The remaining materials are organized as follows. Section \ref{sec:background} and \ref{sec:method} detail the theoretical background and the proposed method. Section \ref{sec:experiments} describes experiment settings. Section \ref{sec:results} and \ref{sec:discussion} present experiment results and discussions.

\section{Background}\label{sec:background}

\subsection{MR Image Reconstruction}
In an MRI system with multiple receiver coils,
the signal from the \(l\)th coil can be modeled as
\begin{equation}
s_l(t)=\int c_l(\vec{r})f(\vec{r})e^{-i2\pi \vec{k}(t)\cdot\vec{r}}\, \mathrm{d}\vec{r},
\label{eq:sl}
\end{equation}
where \(c_l(\vec{r})\) is the coil sensitivity, \(f(\vec{r})\) denotes the finite support continuous function that represents magnetization,
and \(\vec{k}(t)\) is the sampling location in k-space at time \(t\).
During the acquisition, \(s_l(t)\) is sampled at discrete time points,
resulting in the measurements \(\y_l \in \mathbb{C}^M\).
Let \(\x \in \mathbb{C}^N\) represent the discretized form of \(f(\vv{\rr})\),
where $\vv{\rr}\in \mathbb{R}^{N\times d}$
represents the Cartesian grid
of spatial sampling locations.
By stacking the measurements from all \(L\) coils,
we represent the multi-coil forward model \cite{optim_methods_mri} as:
\begin{equation}
\y \defequ
\begin{bmatrix}
\y_1\\
\vdots\\
\y_L
\end{bmatrix}
=
\underbrace{\bm{P}(\bm{I}_L \otimes \bm{F})\,\bm{C}\,}_{\A}\x + \bm{\epsilon}
,\quad
\bm{C} \defequ
\begin{bmatrix}
\bm{C}_1\\
\vdots\\
\bm{C}_L
\end{bmatrix}.
\end{equation}
Here, \(\bm{y} \in \mathbb{C}^{LM}\) is the stacked measurement vector,
\(\bm{F} \in \mathbb{C}^{M\times N}\) represents the discrete Fourier operator,
\(\bm{C}_l \in \mathbb{C}^{N\times N}\) represents the sensitivity of the \(l\)th coil,
and \(\boldsymbol{\epsilon}\in \mathbb{C}^{LM}\) denotes measurement noise.
An optional sampling matrix \(\bm{P}\in \mathbb{C}^{LM\times LM}\) can be included
to account for undersampling k-space.

When the image is fully sampled (\(LM\geq N\)),
the model can be fully determined with proper coil arrangement.
However, to accelerate MRI acquisitions,
fewer k-space samples (\(M < N\)) are acquired,
typically leading to an underdetermined forward operator \A.
Consequently, the naive least-squares solution is non-unique
and introduces undersampling artifacts in images,
necessitating a regularizer \(\mathcal{R}(\x)\) in the image reconstruction problem 
that can be formulated as:
\begin{equation}
\x^* \defequ \argmin_{\x}\|\y-\A\x\|_2^2+\lambda\mathcal{R}(\x).
\label{eq:inverse_problem}
\end{equation}
Common choices for \(\mathcal{R}(\cdot)\) include sparsity- or low-rank-based constraints
\cite{sparse_mri, low_rank_matrix_recovery_cardiac}. 

\subsection{Implicit Neural Representation}

An INR can model the target image \(f(\vv{\rr})\)
as a continuous function with respect to a fixed coordinate grid. 
Instead of using a neural network as an explicit regularizer~\cite{modl}, 
INR leverages a small MLP to learn the parametric representation
\(f_{\bmtheta}(\vec{\rr})\) of the image with respect to the network weights \(\bmtheta\). 
By reformulating the optimization problem in \eqref{eq:inverse_problem} as an optimization over \(\bmtheta\), and denoting \(f(\cdot):\mathbb{R}^{N\times d}\rightarrow \mathbb{R}^{N\times 2}\) as the function that maps coordinates \(\vec{\rr}\in \mathbb{R}^{N\times d}\) to the stacked real and imaginary parts of the image, the self-supervised INR for MRI reconstruction can be formulated as:
\begin{equation}
\label{eq:INR_recon}
\x_{\hat{\bmtheta}} = f_{\hat{\bmtheta}}(\vv{\rr}),\quad
\hat{\bmtheta} = \argmin_{\bmtheta}\|\y-\A f_{\bmtheta}(\vv{\rr})\|_2^2,
\end{equation}
\begin{equation}
\label{eq:INR_function}
f_{\bmtheta}(\vv{\rr})=M_{\bmtheta_{\mathrm{MLP}}}\bigl(\gamma_{\bmtheta_{\mathrm{Enc}}}(\vv{\rr})\bigr),
\end{equation}
where \(\hat{\bmtheta}\) denotes the optimized network parameters under a selected set of 
hyperparameters, 
\(M_{\bmtheta_{\mathrm{MLP}}}\) represents the MLP, and 
\(\gamma_{\bmtheta_{\mathrm{Enc}}}\) is the positional encoding function.
The ``training'' process at reconstruction time
updates the network weights \(\bmtheta\) iteratively
to minimize the data-fidelity term in \eqref{eq:INR_recon}, while the small MLP 
implicitly enforces continuity constraints in the reconstructed image.

\section{Methodology}\label{sec:method}

\subsection{Overall Framework}

\begin{figure*}[t]
\centering
\includegraphics[width=0.95\linewidth]{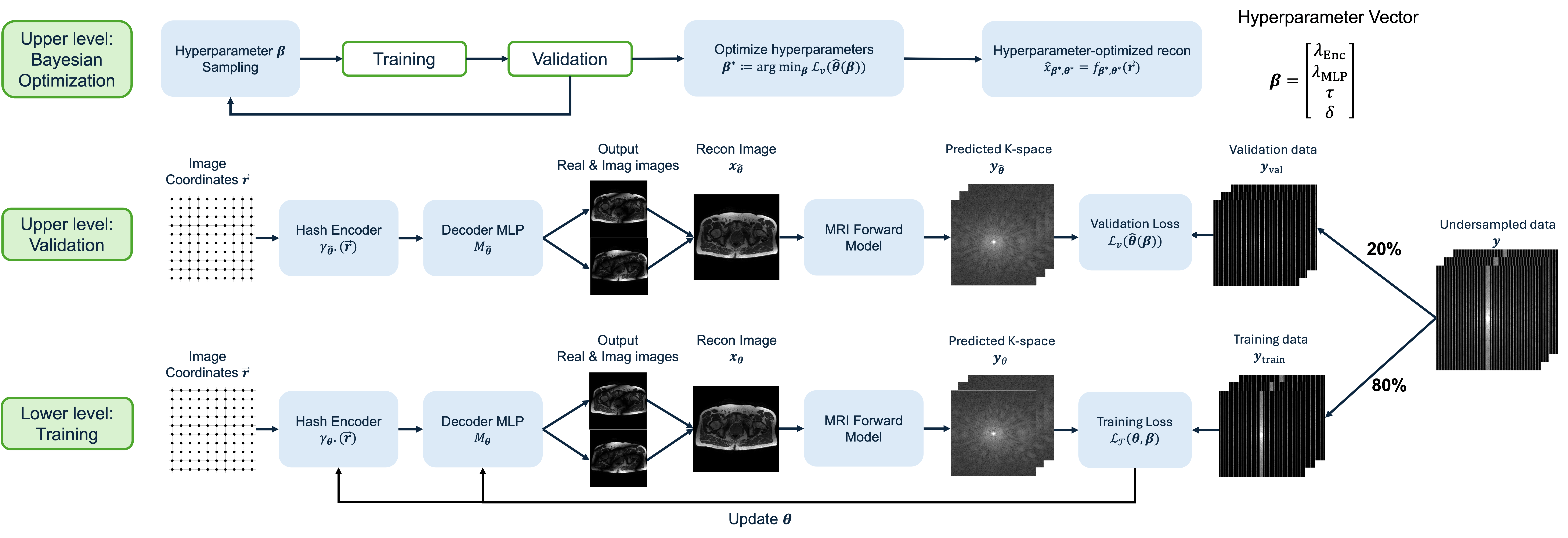}
\caption{
Proposed bilevel-optimized INR framework. The 
undersampled data \(y\) is split into training and validation sets. In each 
upper-level Bayesian optimization iteration, a hyperparameter vector 
\(\bmbeta\) is selected and used to train the INR on \(y_{\textrm{train}}\). 
The trained INR is then validated on \(\y_{\textrm{val}}\) using a weighted 
\(\ell_2\) loss.
The final INR reconstruction and hyperparameter set are 
chosen based on the smallest validation loss.
}
\label{fig:framework}
\end{figure*}

Fig. \ref{fig:framework} shows the proposed bilevel optimized INR framework for MR reconstruction, which has two levels: 
(1) lower-level INR-based image reconstruction, 
and (2) upper-level hyperparameter optimization using the Bayesian Optimization~\cite{bayesopt} with data split from undersampled k-space~\cite{SSDU}. 

We formulate this bilevel optimization problem as:
\begin{equation}
\label{eq:bilevel_inr}
\begin{aligned}
\bmbeta^* 
&\defequ \argmin_{\bmbeta} 
\mathcal{L}_{\mathcal{V}}\bigl(\hat{\bmtheta}(\bmbeta)\bigr) \\
\text{subject to } 
\hat{\bmtheta}(\bmbeta) &\defequ 
\argmin_{\bmtheta} 
\mathcal{L}_{\mathcal{T}}(\bmbeta,\bmtheta),
\end{aligned}
\end{equation}

\noindent where
\begin{equation}
\label{eq:weighted_loss}
\begin{aligned}
\mathcal{L}_{\mathcal{V}}\bigl(\hat{\bmtheta}(\bmbeta)\bigr)
&\defequ
\bigl\|\W\bigl(\y_{\mathrm{val}} - \A f_{\hat{\bmtheta}(\bmbeta)}(\vv{\rr})\bigr)\bigr\|_2^2,\\
\mathcal{L}_{\mathcal{T}}(\bmbeta,\bmtheta)
&\defequ
\bigl\|\W\bigl(\y_{\mathrm{train}} - \A f_{\bmtheta(\bmbeta)}(\vv{\rr})\bigr)\bigr\|_2^2.
\end{aligned}
\end{equation}
Here, \(\mathcal{L}_{\mathcal{T}}(\bmbeta, \bmtheta)\) denotes the lower-level 
training loss and \(\mathcal{L}_{\mathcal{V}}\bigl(\hat{\bmtheta}(\bmbeta)\bigr)\)
is the upper-level validation loss.  
\(\W\in\mathbb{C}^{LM\times LM}\) represents a k-space weighting operator, 
emphasizing the higher frequency components.
\(\bmbeta\) is a  hyperparameter vector,
including \(\ell_2\) regularization strengths 
\(\lambda_{\mathrm{Enc}}\) and \(\lambda_{\mathrm{MLP}}\),
the learning rate \(\tau\), and the loss-weight controller \(\delta\).

Following \cite{SSDU},
we randomly split the undersampled k-space data
into a training set \(\mathcal{T}\)and a validation set \(\mathcal{V}\)
(e.g., an 80\%/20\% split).
In each upper-level iteration,
the algorithm selects a hyperparameter vector \(\bmbeta\)
and trains an INR by iteratively passing the vectorized image coordinates
into a multi-resolution hash encoder 
\(\gamma_{\bmtheta}(\cdot): \mathbb{R}^d \to \mathbb{R}^{LF}\), 
encoding each spatial location into a higher-dimensional feature space. 

An MLP \(M_{\theta_{\mathrm{MLP}}}
(\cdot):\mathbb{R}^{N\times LF}\rightarrow\mathbb{R}^{N\times 2}\) 
decodes these features into the real and imaginary parts 
of the target image \(\x_{\bmtheta}\).
The reconstructed image is then passed through the MR forward model
to produce \(\y_{\bmtheta}\).
The parameters of the encoder and decoder are updated
by calculating a weighted \(\ell_2\) loss
between  \(\y_{\bmtheta}\) and the acquired undersampled k-space \(\y_{\textrm{train}}\).
Once training converges,
the validation loss
compares the predicted k-space
with the unseen measurements \(\y_{\textrm{val}}\).
After a predefined number of upper-level Bayesian optimization iterations,
the method selects
the hyperparameters \(\bmbeta^*\)
and the associated reconstruction \(\hat{\x}_{\bmbeta^*,\bmtheta^*}\) 
having the lowest validation loss.

\subsection{INR for MRI reconstruction}\label{sec:INR_components}

Our INR approach for MRI reconstruction involves three key components:
a positional encoder, a decoder MLP, and the loss function. 
We outline each component below.

\subsubsection{Positional Encoder}
Positional encoding lifts fixed spatial coordinates from \(\mathbb{R}^d\)
to a higher-dimensional feature space \(\mathbb{R}^F\),
allowing a relatively small MLP decoder 
to learn high-frequency variations of the target function. 
In this work, we adopt a multiresolution hash grid encoding \cite{instantngp}
that arranges additional trainable parameters and stores them in an auxiliary data structure,
such as multi-level grids.

\begin{figure}[t]
\centering
\includegraphics[width=\columnwidth]{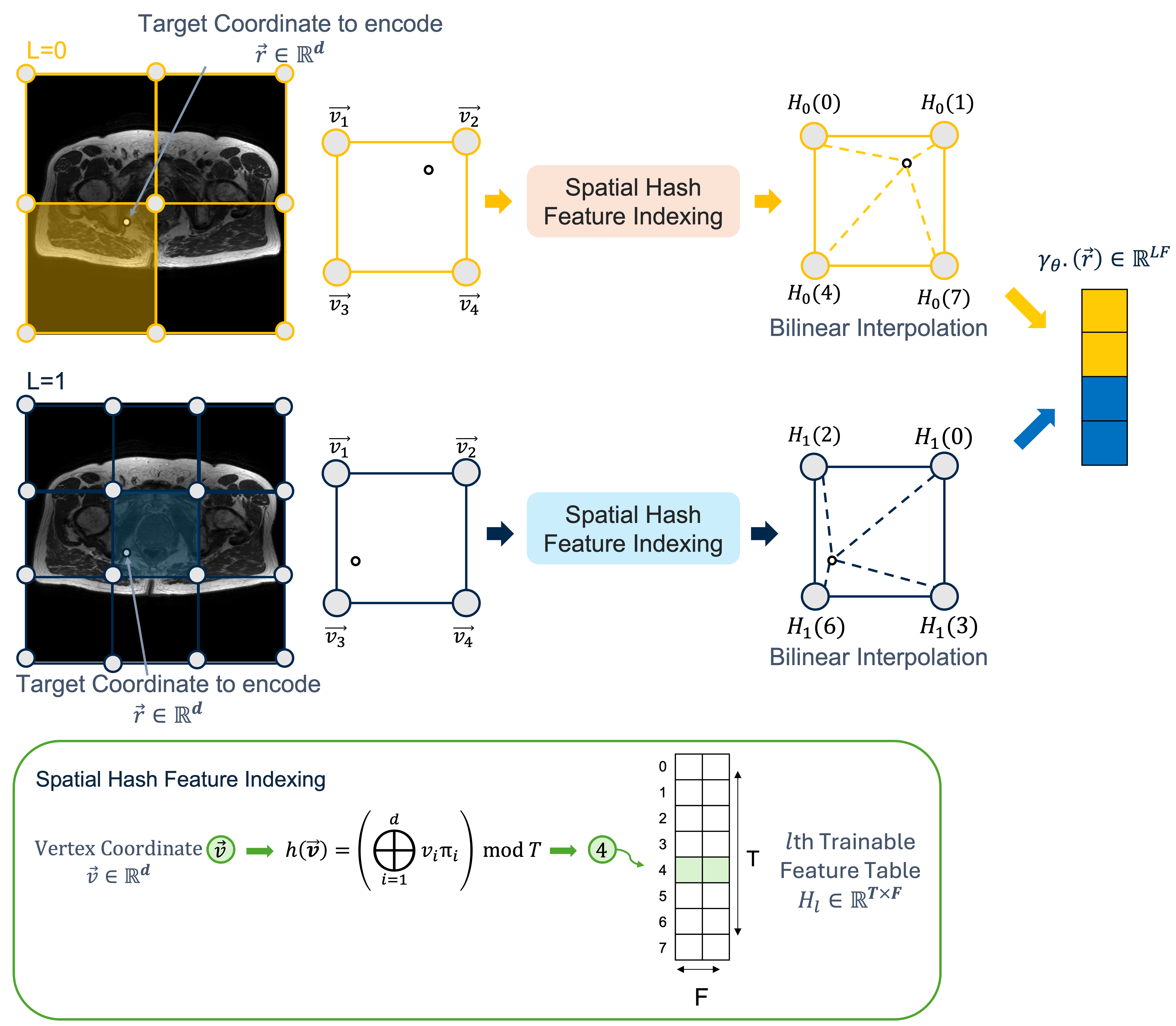}
\caption{Multiresolution Hash Encoding Function \(\gamma_{\theta}.(\vec{r})\)
(e.g., \(\mathnormal{L}=2\))}
\label{fig:hash}
\end{figure}

As illustrated in Fig.~\ref{fig:hash}, the 
target image \(f_{\theta}(\vec{r})\) is represented on \(L\) grid levels (e.g., \(L=16\)), where 
each level is divided into \((a\,b^l) \times (a\,b^l)\) blocks, \(a\) is the size 
of the coarsest grid, and \(b\in(1,2]\) is the scaling factor between levels. 
At each level, the feature for a given coordinate is bilinearly interpolated from 
the features at the vertices of its enclosing block.

The vertex feature \(\vec{v}_i\) is fetched from a trainable parameter table via 
spatial hash indexing, defined by
\begin{equation}
h(\vec{v})=\bigl(\bigoplus_{i=1}^d v_i\,\pi_i\bigr)\bmod T,
\label{eq:spatial_hash}
\end{equation}
where \(\vec{v}\) is the vertex coordinate, \(\pi_i\) is a large prime number, and 
\(T\) is the trainable parameter table size. Stacking all \(L\) levels yields a 
pointwise encoder \(\gamma_{\theta}(\vec{r}):\mathbb{R}^d\rightarrow\mathbb{R}^{LF}\). 

This design captures both high-resolution detail and continuity 
of function manifold through multiresolution grids and interpolation. 
It also speeds up training because only the parameters around the 
queried coordinates are updated, 
unlike an MLP where all parameters are backpropagated.
The size of the trainable hash table is fixed for each level, resulting in hash collision
in fine resolution levels and inducing a stronger gradient that can guide the network
training. 

\subsubsection{Multilayer Perceptron Decoder}
The MLP functions as a decoder that maps the positionally 
encoded features to the target image intensities.
Its learnable function class is determined by its activation function, width and depth.
In this work, we use an 8-layer MLP with 64 neurons per layer and ReLU activation.

\subsubsection{Loss Function}
Due to a large magnitude difference between the lower frequency components at the center and the  
higher frequency components at the periphery of k-space, we propose a modified self-weighting loss that weights each network predicted k-space sample by its magnitude rather than separate real and imaginary channels \cite{spatiotemporal_inr} similar to RGB channels \cite{rawnerf}. The loss function is defined  as:
\begin{equation}
\W = \mathrm{Diag}\bigl(|\A f_{\bmtheta}(\vv{\rr})|+\delta\,\bm{1}\bigr)^{-1},
\label{eq:loss_weight}
\end{equation}
where here the absolute value is applied element-wise.
This self-weighted loss emphasizes higher frequency components and attenuates lower ones, thus avoiding an overly smooth reconstruction. It can also be interpreted as a learnable density compensation 
function that evolves over iterations.
Although no explicit regularizer is applied, the loss function includes \(\ell_2\) regularization terms for the encoder and MLP weights. Their strengths are controlled by \(\lambda_{\mathrm{Enc}}\) and \(\lambda_{\mathrm{MLP}}\), respectively.

\subsection{Bayesian Optimization}
The upper-level hyperparameter optimization in \eqref{eq:bilevel_inr} 
is computationally expensive because each evaluation requires training a new INR. 
Using gradient-based methods would be challenging due to computational complexity,
potential optimization difficulties, and the presence of discrete hyperparameters.
To address these issues, we adopt a zeroth-order method called Bayesian optimization~\cite{bayesopt}
that solves this hyperparameter optimization
by building a Gaussian Process (GP) regression model over the objective function \(f(\bmbeta)\).
After each evaluation of \(\bmbeta\)
(i.e., training an INR and measuring the validation loss),
the GP posterior \(p(f(\bmbeta) \mid \mathcal{D})\) is updated,
where \(\mathcal{D}\) denotes the accumulated data from all previous evaluations.
This surrogate model yields both a predicted mean \(\mu(\bmbeta)\) and uncertainty \(\sigma(\bmbeta)\).
To select the next \(\bmbeta\),
we use the upper confidence bound (UCB) acquisition function
\begin{equation}
    \alpha_{\mathrm{UCB}}(\bmbeta) = \mu(\bmbeta) + \kappa\,\sigma(\bmbeta),
\label{eq:UCB}
\end{equation}
where \(\kappa\) balances the trade-off between exploration (high-\(\sigma\) regions) versus 
exploitation (low predicted loss).
By iteratively updating the GP and evaluating \(\alpha_{\mathrm{UCB}}\),
this method efficiently converges toward near-optimal hyperparameters
with far fewer total network trainings compared to grid or random searches.

\subsection{Implementation Details}

The proposed bilevel optimized INR tunes four hyperparameters in the upper-level 
problem: the learning rate \(\tau\), the loss weighting controller \(\delta\), 
and the weight decay terms \(\lambda_{\mathrm{Enc}}\) and \(\lambda_{\mathrm{MLP}}\)
for the encoder and MLP, respectively.
We used 60 upper-level iterations,
each consisting of 4000 lower-level iterations,
to ensure convergence of the INR reconstruction. 
All image coordinates were normalized to \((0,1)\) in each dimension.

The Hash encoder included 16 levels with \(2^{18}\) parameters per level for 2D image reconstruction.
The decoder MLP had six hidden layers, each with 64 ReLU neurons,
while the input and output layers used no activation.

Coil maps were estimated using E-SPIRiT~\cite{ESPIRIT} from a separate gradient-echo 
(GRE) scan or from fully sampled center k-space lines.
Virtual coil compression~\cite{coil_compress} 
and noise prewhitening~\cite{noise_prewhiten} were applied before reconstruction. 
Nonuniform fast Fourier Transforms~\cite{NUFFT} for non-Cartesian acquisitions
were implemented via the \texttt{torchkbnufft} toolbox~\cite{kbnufft}.

The study was implemented in PyTorch 1.12.1 (Python 3.9) on a high-performance computing cluster
using an NVIDIA A40 GPU and a 2.9\,GHz Intel Xeon Gold 6226R CPU with 16\,GB of memory. 
The source code is available at \url{https://github.com/MIITT-MRI-Jianglab/Bilevel_optim_INR}.

\section{Experiments}\label{sec:experiments}

\subsection{Experiments}

\subsubsection{Comparison with prior art}

We compared the image quality of bilevel optimized INR with several other model-based and self-supervised reconstruction methods,
such as CG-SENSE~\cite{CGSENSE}, \(\ell_1\)-wavelet~\cite{sparse_mri}, P-LORAKS~\cite{PLORAKS}, and IMJENSE~\cite{IMJENSE}. 
The comparisons were performed using data acquired across various anatomies, sampling patterns, and 
field strengths. Specifically, we used balanced SSFP (bSSFP) and T2-weighted turbo spin echo (TSE) datasets to compare Cartesian and Poisson undersampling patterns. Additionally, we evaluated the 
reconstruction results of a healthy cardiac volunteer using a T2-weighted bSSFP 
sequence on a 0.55T scanner. To further demonstrate capabilities of our technique
for non-Cartesian sampling, 
we simulated a spiral acquisition from a fully sampled T2-weighted TSE prostate 
dataset at 3T. The spiral is designed using minimum-time gradient 
design \cite{mintime_gradient} with zero-moment compensation, and requires
48 interleaves to fully sample the k-space.

\subsubsection{Ablation Study of INR Design}

We performed experiments to determine the most effective formulation of 
INR for MRI reconstruction and validate the choices in Section~\ref{sec:INR_components}. 
We examined the impact of different decoders, such as varying 
numbers of MLP layers and using a linear layer, as well as different positional encoders, including no 
encoder, frequency encoding \cite{nerf}, dense grid encoding \cite{dense_grid}, and hash encoding \cite{instantngp}. For the loss function comparison, we tested several weighting strategies, such as identity weighting, density 
compensation function weighting \cite{jpipe_dcf}, acquired data weighting, and self-weighting as described in~\eqref{eq:loss_weight}.  We also conducted experiments on activation function, comparing ReLU and Sinusoidal activation~\cite{siren}.

\subsubsection{Hyperparameter Optimization}

We examined three hypotheses on hyperparameter optimization in bilevel-optimized INR for image reconstruction. (H1: Comparable Quality to an Oracle) Our bilevel approach is hypothesized to yield reconstructions comparable to an oracle that selects near-optimal hyperparameters using fully sampled data for validation loss; we evaluated H1 by comparing our reconstructions against this oracle standard using identical datasets and acquisition parameters. (H2: Protocol-Specific Optimization) We tested whether hyperparameter tuning is necessary for different imaging protocols by applying hyperparameters optimized for bSSFP to T2-weighted TSE data from the same subject, assessing performance consistency. (H3: Transferability Across Subjects/Anatomies) We hypothesized that hyperparameters optimized for one subject or anatomy can be effectively transferred to another under the same protocol; to test H3, we first conducted bSSFP acquisitions on two volunteers, comparing reconstructions using each subject’s optimized hyperparameters and those from the other, and then acquired multislice prostate T2w imaging to perform a similar comparison across different slices.

\subsubsection{Real-time interventional MRI}

We validated the proposed algorithm's capability for real-time 
residual learning and hyperparameter transferability for the same anatomy
using the MR-guided biopsy liver phantom dataset.
We used the fully sampled first time frame to optimize the hyperparameters and pre-train the INR. 
We then retrospectively \(6\times\) undersampled subsequent time frames with \(4\%\) autocalibration region signal (ACS) lines and fine-tuned the INR for 500 iterations for each frame.

Table \ref{tab:acquisition_parameters}
summarizes the acquisition parameters of the above experiments. The temporal resolution for the iMRI is 2 seconds per frame.
\begin{table}[t]
\centering
\caption{Acquisition Protocols and Parameters}
\label{tab:acquisition_parameters}
\resizebox{\columnwidth}{!}{%
\begin{tabular}{lcccccc}
\toprule
\textbf{Imaging Type}  & \textbf{Coil No.} & \textbf{Sequence} & \textbf{Matrix Size} & \textbf{Resolution (mm\(^2\) / mm\(^3\))} \\ 
\midrule
Brain (3T) & 16 & bSSFP / T2w TSE & $256 \times 256$ & $1 \times 1$ \\ 
Cardiac (0.55T) & 15 & T2w bSSFP & $294 \times 272$ & $1.4 \times 1.4$ \\ 
Prostate (3T) & 12 & T2w TSE & $384 \times 384 \times 32$ & $1 \times 1 \times 3$ \\ 
iMRI (0.55T) & 24 & bSSFP & $26 \times 128 \times 128$ & $2.3 \times 2.3$  \\ 
\bottomrule
\end{tabular}%
}
\end{table}

\subsection{Performance Measurement}

Performance in this study was quantified using three metrics: Normalized 
Root-mean-squared Error (NRMSE), Structural Similarity Index (SSIM), and 
Peak Signal-to-noise Ratio (PSNR). Reconstructions were compared against the 
ground truth from fully sampled data in both retrospective and simulated 
studies. For all metrics calculations, the coil map estimated by E-SPIRiT 
was used as the region-of-interest (ROI) mask and code implementation is 
based on the scikit-image package\footnote{https://scikit-image.org/}.

\section{Results}\label{sec:results}

\subsection{Comparisons with other Self-supervised Reconstruction methods}

\begin{figure*}[ht]
\centering
\includegraphics[width=0.9\linewidth]{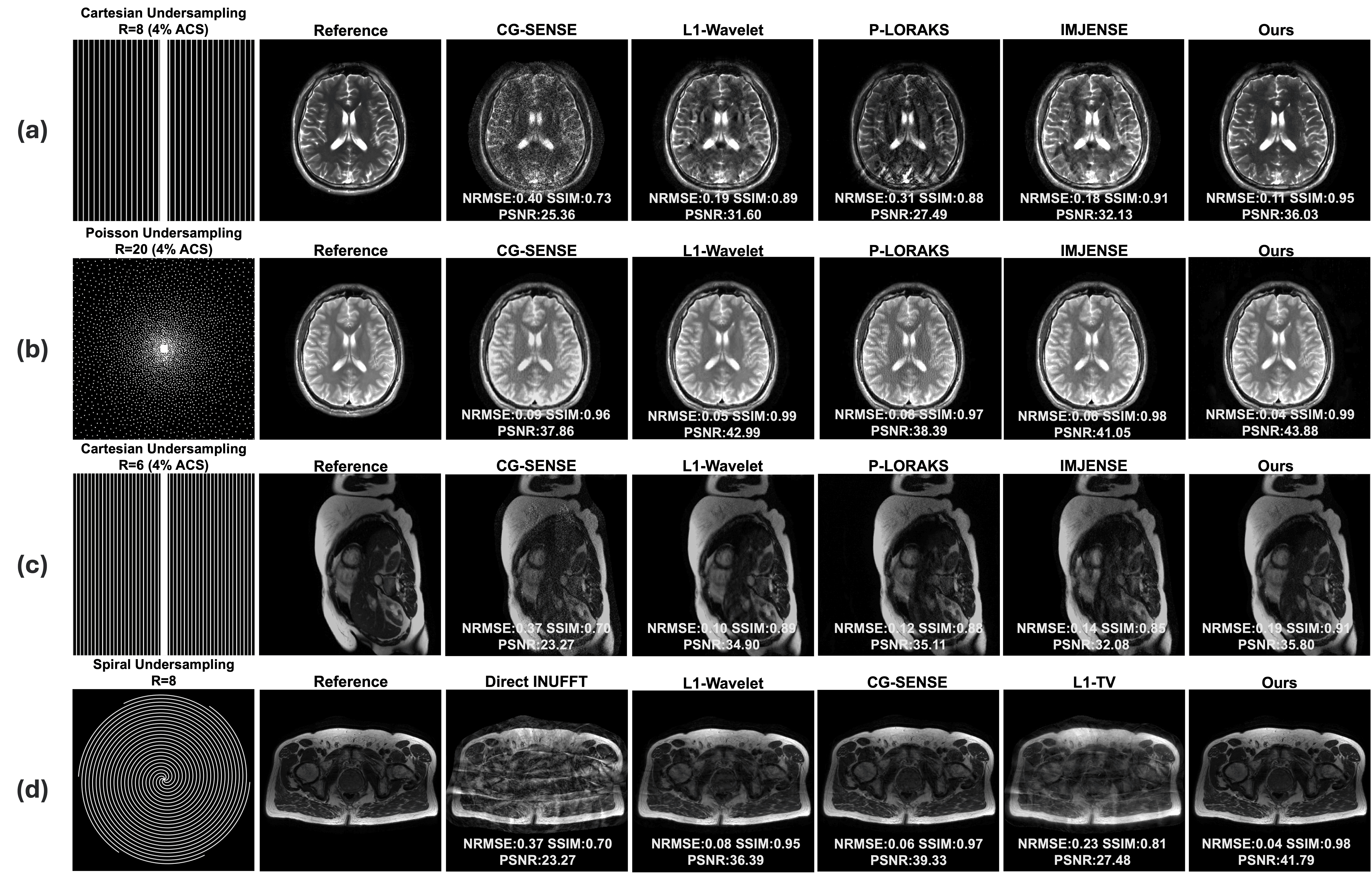}
\caption{Method comparisons.
(a) Brain volunteer at 3T using bSSFP of \(1\times1\ \textrm{mm}^2\) resolution.
(b) Brain volunteer at 3T using T2w TSE of \(1\times1\ \textrm{mm}^2\) 
(c) Cardiac volunteer at 0.55T using T2w bSSFP of \(1.4\times1.4\ \textrm{mm}^2\)
resolution.
(d) Prostate volunteer at 3T using T2w TSE of \(1\times1\ \textrm{mm}^2\) resolution.
Detailed acquisition protocols are shown in Table~\ref{tab:acquisition_parameters}.
}
\label{fig:method_comparison}
\end{figure*}

Fig.~\ref{fig:method_comparison} compares our proposed approach with 
other model-based iterative and self-supervised deep-learning methods.

For \(8\times\) Cartesian undersampled bSSFP brain experiments 
(Fig.~\ref{fig:method_comparison}(a)), CG-SENSE introduces notable noise 
across the brain, and \(\ell_1\)-Wavelet, P-LORAKS, and IMJENSE exhibit 
residual artifacts from the sampling pattern. In contrast, the proposed 
method shows only minor residual artifacts and achieves a PSNR more than 
5\,dB higher than the other approaches.

Fig.~\ref{fig:method_comparison}(b) presents results from \(20\times\) Poisson undersampled 
experiments with a central \(10\times 10\) pixel auto-calibration region. 
Both CG-SENSE and P-LORAKS show visible non-structured noise in the whole 
brain. In contrast, \(\ell_1\)-wavelet, IMJENSE, and the proposed bilevel 
INR display visually artifact-free reconstructions, with our method 
outperforming others supported by all metrics.
INR-based methods (i.e., IMJENSE and ours) benefit from the relatively 
uniform sampling pattern, which results in non-structural artifacts. 
These artifacts are easier to remove by INR, because of 
its strength in learning continuous and smooth function manifolds.

Fig.~\ref{fig:method_comparison}(c) compares methods on a non-gated cardiac dataset at 0.55T, 
acquired via a T2-weighted bSSFP sequence. The proposed bilevel INR 
reduces noise and suppresses most motion artifacts by its implicit regularization
on continuity. 
Other methods either 
fail to remove noise (CG-SENSE, IMJENSE) or retain motion artifacts 
(P-LORAKS, IMJENSE). The \(\ell_1\)-Wavelet result appears overly smooth.

Finally, for simulated \(8\times\) spiral undersampling 
(Fig.~\ref{fig:method_comparison}(d)), all model-based methods show spiral undersampling 
artifacts. Our method achieves better metrics and yields nearly 
artifact-free reconstructions. IMJENSE is omitted here because it only 
supports Cartesian sampling patterns.

\subsection{Methods ablation study}
\begin{figure}[h]
\centering
\includegraphics[width=\columnwidth]{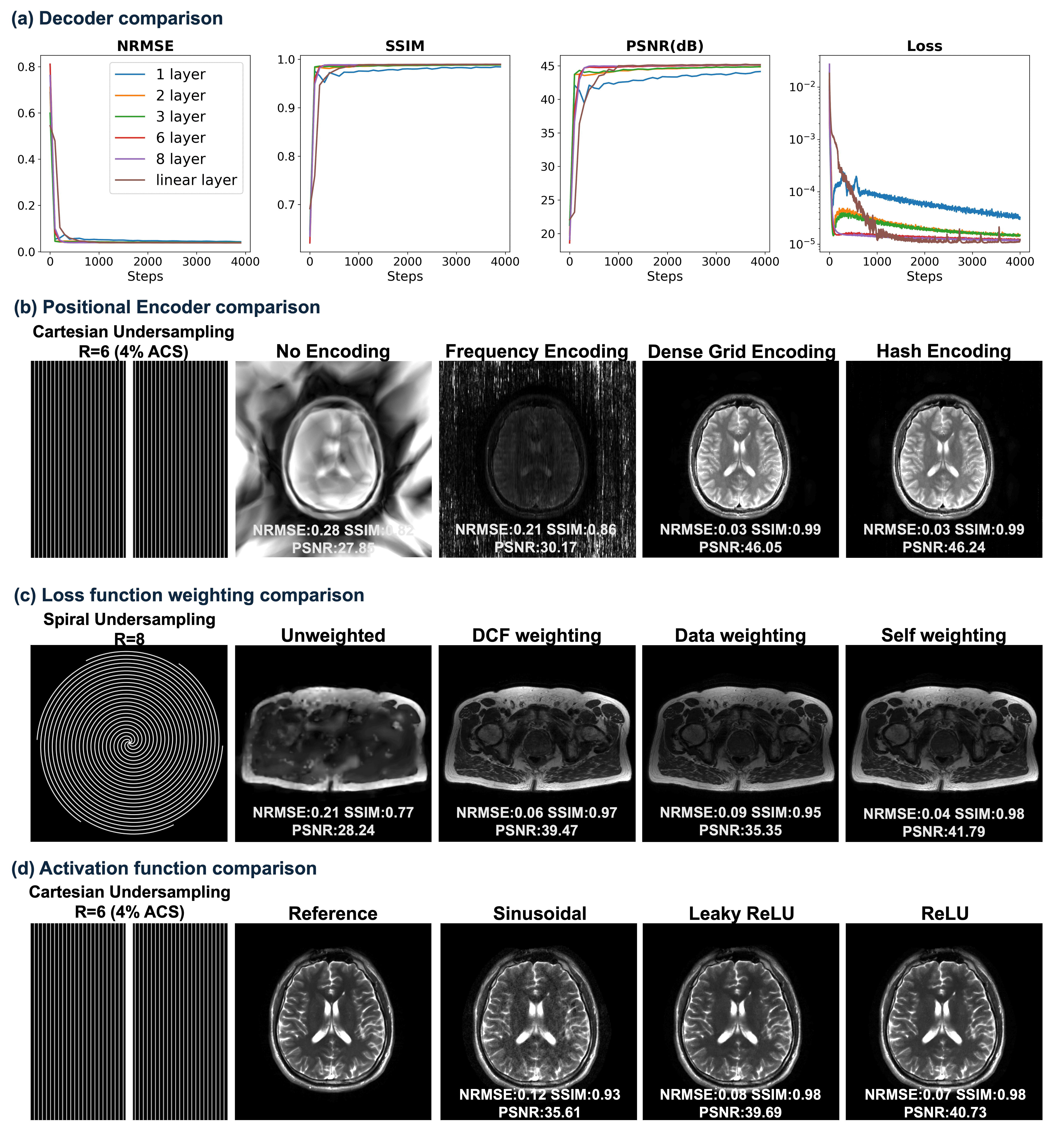}
\caption{Ablation study.
(a) Decoder comparison with fixed Hash Encoder (i.e., Linear layer vs 
1/2/3/6/8 layer MLP).
(b) Encoder comparison with fixed Decoder MLP.
(c) Loss function weighting comparison.
(d) Activation function comparison.
}
\label{fig:ablation}
\end{figure}

Fig.~\ref{fig:ablation} presents ablation experiments on key components 
of the bilevel-optimized INR. Fig.~\ref{fig:ablation}(a) compares decoders of varying 
depths against a linear layer under identical hyperparameters. Except for 
the single-layer MLP, all decoders converge to similar performance based 
on quantitative metrics. Even a linear layer can decode the positionally 
encoded features effectively, indicating the encoder’s dominant role in 
the reconstruction pipeline.

Fig.~\ref{fig:ablation}(b) evaluates different positional encoders on a \(6\times\) 
Cartesian undersampled T2-weighted TSE dataset. Without encoding or with 
frequency encoding, the model struggles to capture fine details in the 
brain, while dense grid and hash grid encodings yield comparable metrics. 
However, dense grid requires \(20\times\) memory and \(15\times\) of training time

Fig.~\ref{fig:ablation}(c) focuses on loss-function weighting, tested on a spiral 
undersampled prostate dataset. The proposed self-weighting scheme in 
\eqref{eq:loss_weight} has the best reconstruction quality, 
capturing correct contrast and detailed structure. No loss 
weighting causes visible blurriness across the pelvic region. Using 
Pipe’s density compensation function~\cite{jpipe_dcf} ranks second, with 
\(0.06\)~NRMSE, \(0.97\)~SSIM, and \(39.47\)~dB~PSNR compared to \(0.04\), 
\(0.98\), and \(41.79\)~dB from our approach. Substituting the acquired 
data for the predicted k-space in each iteration leads to a bad
local minimum during training, which shows incorrect contrast and 
degraded metrics.

\subsection{Hyperparameter optimization and transferability experiments}

\begin{figure*}[ht]
\centering
\includegraphics[width=0.6\linewidth]{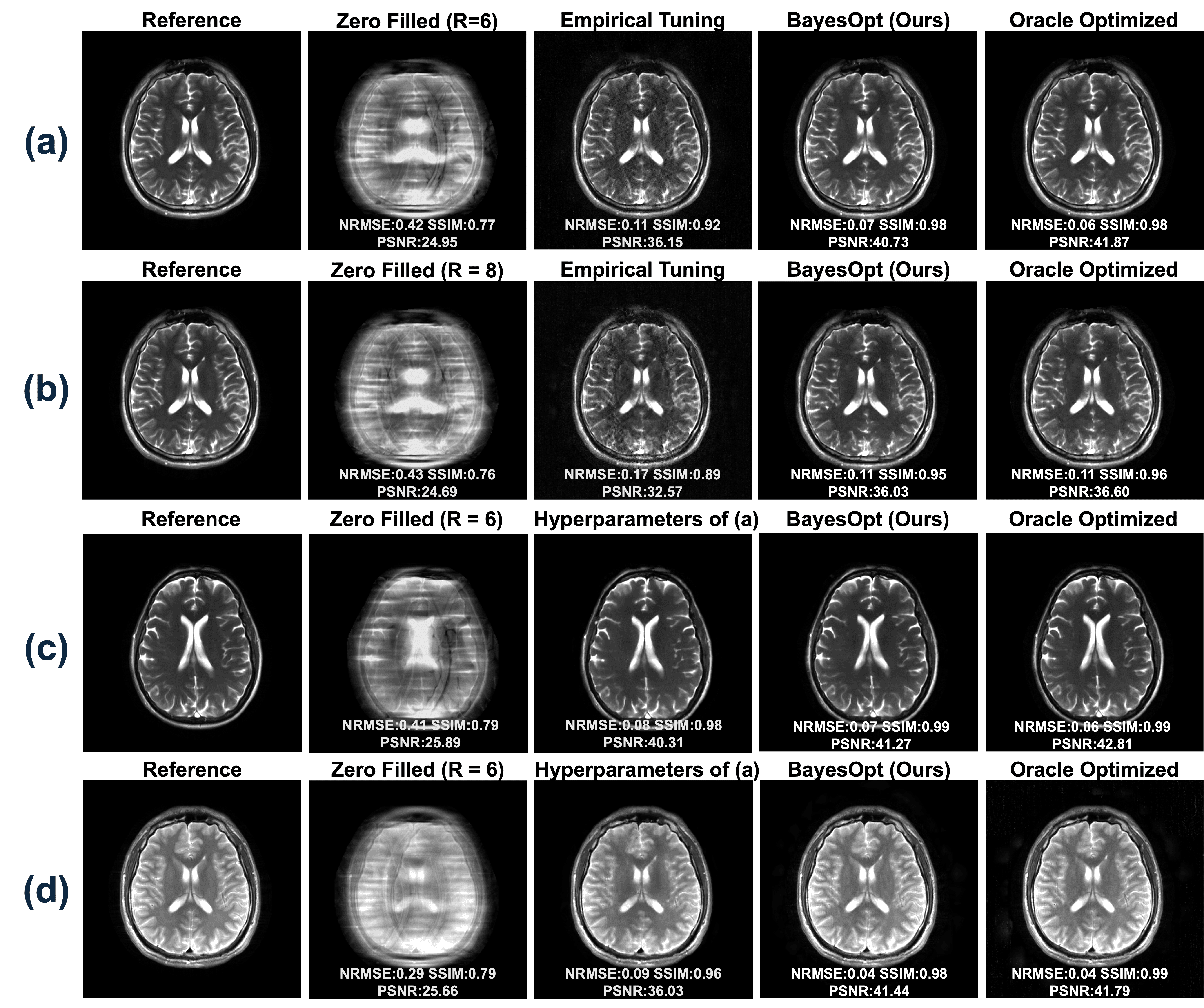}
\caption{
Demonstration of hyperparameter optimization using Bayesian Optimization.
(a) and (b) show comparisons of bSSFP acquisitions at acceleration factors \(R=6\) and \(R=8\) for volunteer~1.
(c) shows comparison of bSSFP acquisition for volunteer~2 at \(R=6\), illustrating that hyperparameters are transferable to a different subject for the same imaging sequence.
(d) shows a comparison of a T2w TSE acquisition from volunteer~1 in (a) at \(R=6\), illustrating that different sequences require tailored hyperparameter optimization.
Acquisition parameters are provided in Table~\ref{tab:acquisition_parameters}.}
\label{fig:bilevel_recon_demo}
\end{figure*}

\begin{figure}[h]
\centering
\includegraphics[width=\columnwidth]{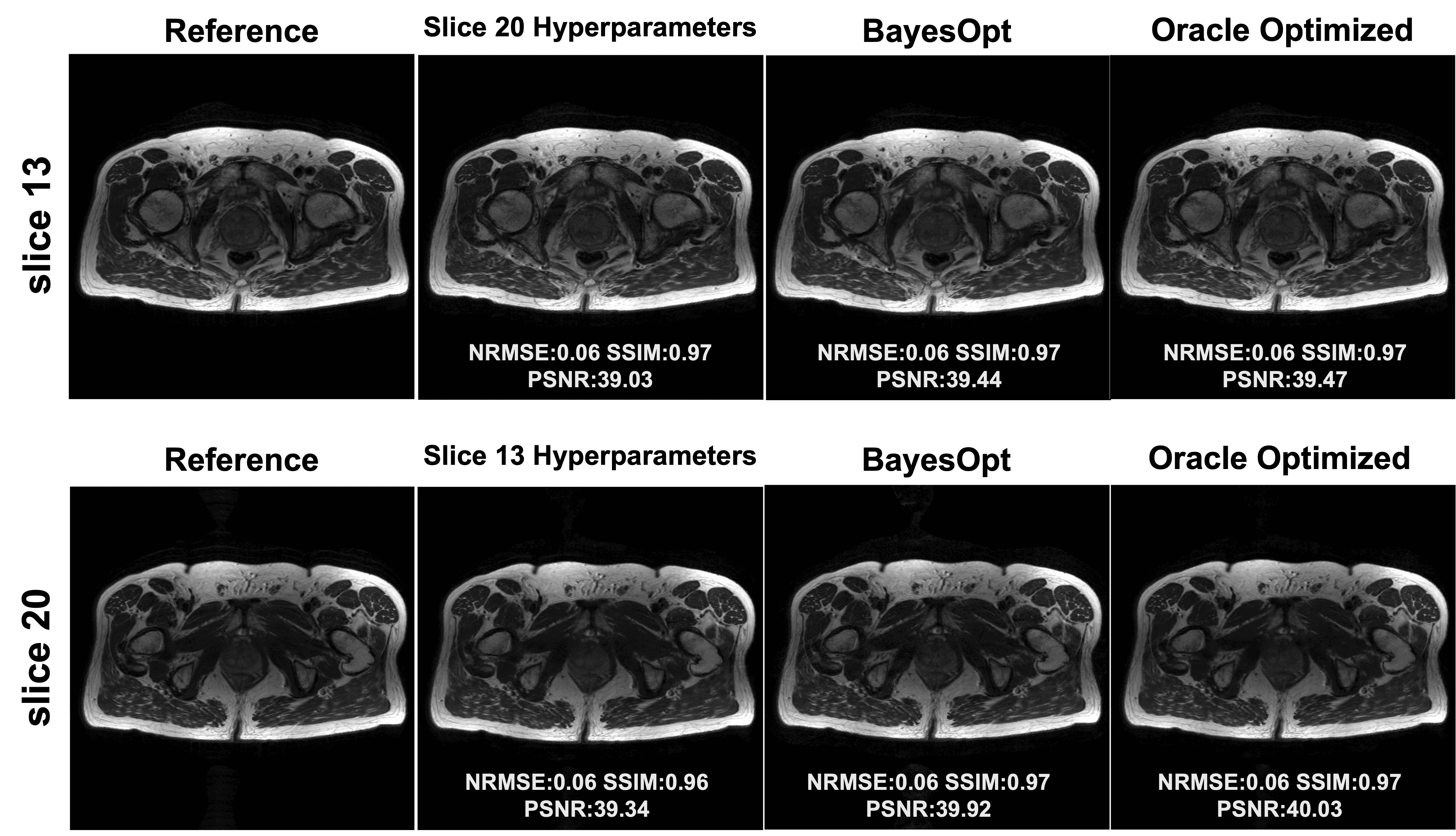}
\caption{
Demonstration of hyperparameter transferability. Reconstructions of two slices from a multislice prostate acquisition are shown (\(6\times\) Cartesian undersampled). Each slice is reconstructed with both its own optimized hyperparameters and those from the other slice, indicating that the optimized hyperparameters are transferable across similar anatomy under the same acquisition.}
\label{fig:transferability}
\end{figure}

\begin{table}[!t]
\caption{Comparison of hyperparameter values found by empirical tuning vs.\ Bayesian optimization.
(Fig.~\ref{fig:bilevel_recon_demo} (a),(b))}
\label{tab:hyperparams}
\centering
\resizebox{\columnwidth}{!}{%
\begin{tabular}{lcccc}
\toprule
 & $\lambda_{\mathrm{ENC}}$ & $\lambda_{\mathrm{MLP}}$ & lr & $\delta$ \\
\midrule
Empirical Tuning & $1.0\times10^{-5}$ & $1.0\times10^{-10}$ & $1.0\times10^{-3}$ & $1\times10^{-4}$ \\
BayesOpt (R=6)   & $2.6\times10^{-4}$ & $1.0\times10^{-7}$  & $9.3\times10^{-3}$ & $2\times10^{-4}$ \\
BayesOpt (R=8)   & $3.6\times10^{-3}$ & $7.2\times10^{-6}$  & $4.4\times10^{-4}$ & $2\times10^{-4}$ \\
\bottomrule
\end{tabular}%
}
\end{table}

Fig.~\ref{fig:bilevel_recon_demo} illustrates
the hyperparameter optimization results. 
Fig.~\ref{fig:bilevel_recon_demo}(a) and (b) show that Bayesian 
optimization yields nearly optimal results compared to the oracle 
optimization across NRMSE, SSIM, and PSNR for both \(R=6\) and \(R=8\) on volunteer~1. 
By contrast, empirical tuning leads to residual artifacts near the brain 
center. The empirically tuned hyperparameters were chosen based on 
previously successful reconstructions for other sequences and resolutions. 
Table~\ref{tab:hyperparams} lists the hyperparameter values in each case. 
Except for the loss-weighting controller \(\delta\), all other 
hyperparameters differ between \(R=6\) and \(R=8\), indicating that each 
sampling pattern benefits from a tailored optimization.

Fig.~\ref{fig:bilevel_recon_demo}(c) displays the reconstruction of the same bSSFP sequence with 
identical acquisition parameters and similar slice position for volunteer~2. 
Bayesian optimization achieves performance comparable to oracle optimization 
across all metrics. The hyperparameters optimized for volunteer~1 adapt well 
to volunteer~2, with similar NRMSE and SSIM, and only slightly worse PSNR, 
showing that hyperparameters optimized for bilevel INR can be transferred 
between similar acquisitions.

Fig.~\ref{fig:bilevel_recon_demo}(d) uses a T2-weighted TSE scan from volunteer~1 at the same slice 
position to demonstrate the sequence dependence of INR-based reconstruction.
The performance using hyperparameters optimized for the 
bSSFP scan of volunteer~1 is \(5.41\)~dB worse in PSNR and \(0.05\) higher in 
NRMSE compared to tailored Bayesian Optimization. The resulting images appear 
more blurry because bSSFP sequences generally have lower SNR than TSE, 
and optimizing for bSSFP tends toward stronger denoising (e.g., \(\lambda_{\mathrm{Enc}}=2.6e-4\) in (a)). 
In contrast, T2-weighted TSE scans keep anatomical structures more clearly 
and thus need less regularization (e.g., \(\lambda_{\mathrm{Enc}}=1.2e-5\) in (d)) .

Fig.~\ref{fig:transferability} further demonstrates the transferability of 
hyperparameters across anatomically similar regions. 
Reconstructions for two \(6\times\) undersampled pelvic slices are compared using 
both tailored hyperparameters (optimized specifically for each slice) 
and hyperparameters transferred from the other slice. 
The transferred hyperparameters yield good reconstructions, 
with only small decreases in PSNR (0.41 dB and 0.58 dB, respectively) compared to the tailored results.

\subsection{Examples of tuned hyperparameters}

\begin{figure}[h]
\centering
\includegraphics[width=\columnwidth]{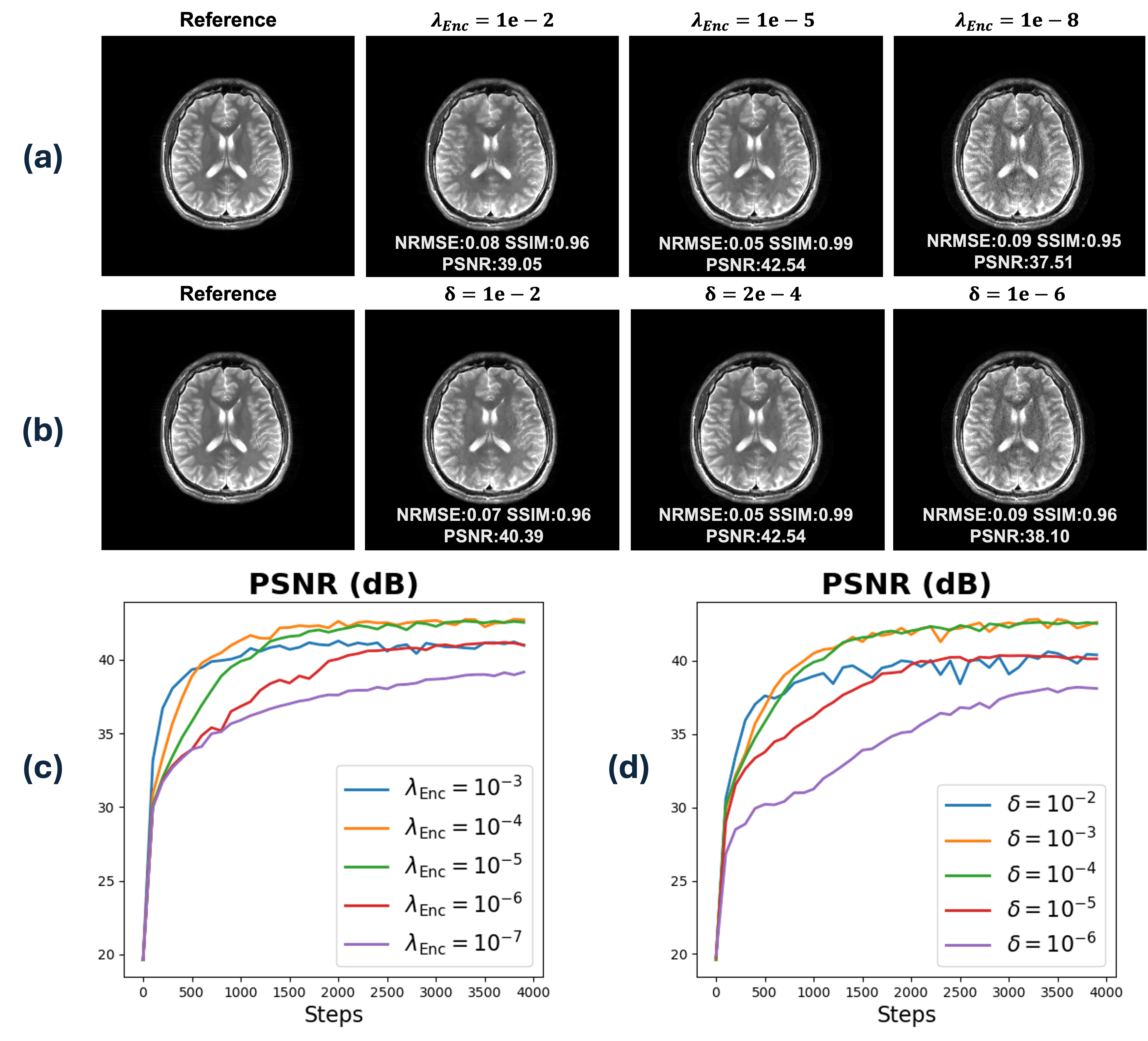}
\caption{Illustrating choosing hyperparameter for Bayesian Optimization;
(a),(c) \(\ell_2\) regularization strength for Hash Encoder parameters.
(b),(d) Self-weighting loss stability value \(\delta\) that controls 
the emphasis on higher frequency k-space components.
}
\label{fig:hyperparameters}
\end{figure}

Fig.~\ref{fig:hyperparameters} demonstrates the impact of optimized 
hyperparameters in the upper level of \eqref{eq:bilevel_inr}. The 
encoder’s \(\ell_2\) regularization strength \(\lambda_{\mathrm{Enc}}\) 
adjusts the degree of regularization on the positionally encoded 
feature representation; sub-optimal choices lead to over- or 
under-regularization. Similarly, the loss-weighting controller 
\(\delta\) balances the trade-off between image smoothness and fine 
details. According to \eqref{eq:loss_weight}, a larger \(\delta\) 
diminishes higher-frequency components relative to the lower-frequency 
terms, resulting in blurrier reconstructions. Conversely, smaller 
\(\delta\) emphasizes high-frequency components, producing clearer 
but potentially more aliased images.

\subsection{Real-time iMRI}

\begin{figure}[h]
\centering
\includegraphics[width=\columnwidth]{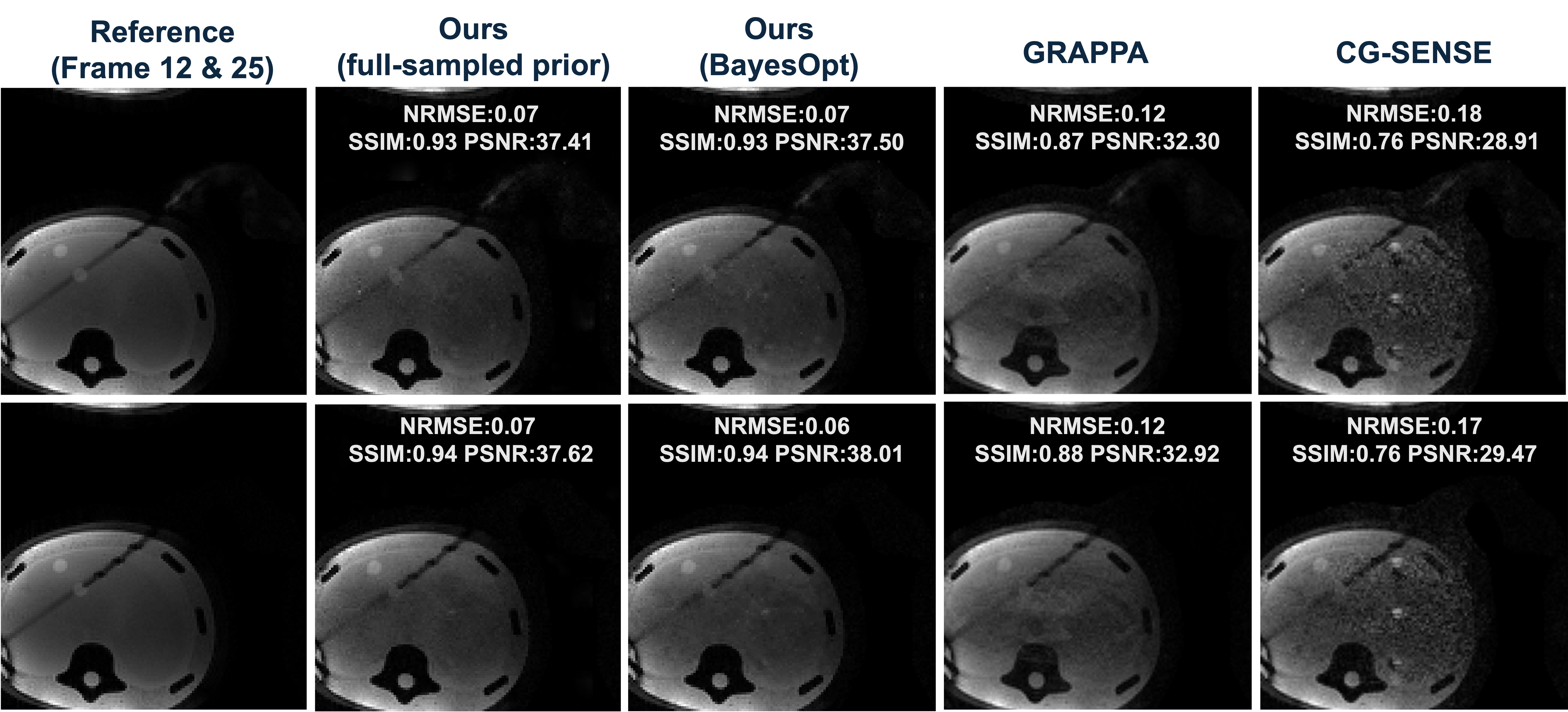}
\caption{
Comparisons of real-time iMRI reconstructions. The INR is pre-trained on the fully sampled first frame. Each subsequent frame is \(6\times\) undersampled with \(6\) ACS lines and reconstructed using only \(500\) residual-learning iterations. Reconstructions of two selected frames obtained using our method with the fully sampled prior and frame-specific tailored reconstructions without prior are compared with those from GRAPPA and CG-SENSE. Acquisition parameters are provided in Table~\ref{tab:acquisition_parameters}.
}
\label{fig:iMRI}
\end{figure}

Fig.~\ref{fig:iMRI} compares real-time interventional MRI reconstructions of a liver phantom. The first frame was fully sampled to train the INR, while subsequent frames were retrospectively undersampled by \(6\times\). Our INR-based residual learning reconstruction outperforms GRAPPA and CG-SENSE by over 5\,dB in PSNR, with fewer visible artifacts and noise. The needle and target biopsy region remain clearly distinguishable. Because the INR is pre-trained on the first frame, reconstructing each subsequent frame takes only 1\,s (\(<500\) iterations), which is faster than the 2\,s temporal resolution required for this real-time acquisition.
We also compared this residual learning scheme with frame-specific tailored reconstruction, in which the complete bilevel optimization is performed for each frame. The residual learning reconstruction achieves similar quantitative metrics. A movie showing the full time-series reconstruction, including comparisons at different acceleration factors (\(R = 4/6\)), is provided in the supplementary material.

\section{Discussion}\label{sec:discussion}

This paper proposes a bilevel-optimized INR 
framework for hyperparameter-optimized, scan-specific MRI reconstruction for 
accelerated MRI acquisitions.
Existing self-supervised deep learning methods rely on 
task- and subject-dependent hyperparameter tuning~\cite{IMJENSE, NeRP, DNLINV},
which can be impractical in clinical settings. In contrast, we formulate 
the MRI reconstruction as a bilevel optimization problem, where the upper level 
optimizes hyperparameters and the lower level performs INR-based reconstruction 
given sampled hyperparameters. 
This approach automatically determines protocol-specific hyperparameters, 
providing improved and tailored reconstruction quality.

We validated the proposed method on \textit{in vivo} scans of healthy volunteers 
across various anatomies, contrasts, sampling patterns, and field strengths. 
Our bilevel-optimized INR achieved higher quantitative metrics (NRMSE, SSIM, and PSNR) 
than both model-based and self-supervised INR methods. 
Furthermore, we showed that the optimized hyperparameters can be transferred to 
other subjects or similar anatomies scanned under the same protocol.

\subsection{Analysis of self-supervised bilevel optimization}

The primary challenge of supervised deep learning methods is the need for large, 
application-specific training datasets and their limited generalization to unseen 
acquisitions. Self-supervised deep learning alleviates this issue by performing 
scan-specific reconstruction. However, existing self-supervised methods require 
dedicated hyperparameter optimization for each dataset.
Current approaches either tune empirically~\cite{DNLINV, NeRP}
or use population-based hyperparameter optimization~\cite{IMJENSE}.

This study addresses these limitations by combining automatic hyperparameter 
optimization and scan-specific self-supervised reconstruction. Instead of 
using population-based hyperparameter optimization, we split the undersampled 
data into further undersampled training and validation sets, inspired by SSDU 
\cite{SSDU}, to achieve self-supervision. 
We then employ Bayesian Optimization for the upper-level search because 
of its efficiency when evaluating multiple hyperparameters in an expensive lower-level 
INR-based image reconstruction (i.e., training an INR). Bayesian Optimization also 
handles discrete inputs and only requires a fixed number of iterations (e.g., 20 for 
initialization and 40 for Gaussian regression).

We focus on four hyperparameters that are highly application-dependent: the weight 
decay for encoder and decoder parameters, the learning rate, and the loss-weighting 
controller. Both weight decay parameters balance the trade-off between noise robustness and structural detail, where \(\lambda_{\mathrm{Enc}}\) regularizes positionally encoded latent features and \(\lambda_{\mathrm{MLP}}\) directly regularizes the learned function
manifolds. \(\lambda_{\mathrm{Enc}}\) dominates due to the hash encoder's higher trainable
weight count compared to the compact MLP.
As shown in Fig.~\ref{fig:hyperparameters}, the loss-weighting controller further 
optimizes the balance between blurriness and resolution in the reconstruction. 
We incorporate the learning rate into hyperparameter optimization
because it jointly influences reconstruction results with the other three parameters,
and its optimal value is also interdependent on the others.

We also considered additional hyperparameters such as the MLP’s width and depth, 
but observed that their influence remains relatively stable across different datasets. 
Fig.~\ref{fig:ablation} shows that even a linear-layer decoder can produce high-quality 
reconstructions, demonstrating the positional encoder’s dominant contribution for 
the reconstruction. 
For the Hash encoder, we tested a table size of \(\geq 2^{16}\) and a finest 
resolution determined by scaling factor \(b\geq1.5\), both of which yielded 
comparable performance across various scans. 
Overall, over-parameterization with optimized self-regularization provided by weight decay is another reason
why the hash-encoded INR-based approach can reconstruct highly accelerated MRI scans.

\subsection{Analysis of components of Bilevel Optimized INR}

This study also provides experiments to analyze the effect and selection of key 
components in an INR for scan-specific MRI 
reconstruction. The main components of an INR are the positional encoder, the 
decoder MLP, and the loss function.
Fig.~\ref{fig:ablation}(b) compares no encoding, frequency encoding 
\cite{nerf}, dense grid encoding \cite{dense_grid}, and hash grid encoding \cite{instantngp}. 
Unlike view synthesis, which leverages multiple view images for training, no 
encoding and frequency encoding are both slow to train and even unable to capture 
fine MR image details. By contrast, dense grid and hash grid encodings can 
reconstruct images effectively due to over-parameterization. However, 
the dense grid demands more memory and training time since it 
stores weights for every grid corner.
Fig.~\ref{fig:ablation}(a) compares MLPs of various depths against a linear 
layer. All decoders achieve similar performance upon convergence, consistent 
with prior findings \cite{instantngp} that the encoder primarily solves the 
inverse problem. However, our activation function differs from SIREN \cite{siren}, 
which applies sinusoidal activation for view synthesis.
In this study, ReLU activation yielded more stable training
and was less prone to overfitting,
whereas SIREN tended to introduce residual artifacts.
We attribute this to:
(1) this task is heavily over-parametrized as only a 2D MR image is
reconstructed instead of full 3D view like in SIREN,
and (2) the implicit regularization from the ``dying ReLU'' phenomenon \cite{dying_relu},
which helps limit overfitting.
This also explains why previous Hash-encoded INR MRI reconstruction 
methods \cite{IMJENSE, CEST_INR}
often required additional explicit priors, as sinusoidal activation was used.
We interpret the loss weighting as a learned density compensation function (DCF). 
In non-Cartesian reconstructions, DCF emphasizes the higher-frequency components 
(much smaller in magnitude compared to low-frequency components) 
to avoid image blurring. A similar rationale applies to INR-based reconstruction: 
a standard \(\ell_2\) loss in k-space could lead the MLP to learn a conditional mean
of the target \(k\) space distribution in the regression task\cite{PR_ML}. 
Our experiments show that a learned self-weighting scheme provides the best performance,
displayed in Fig.~\ref{fig:ablation}(c).

\subsection{Limitations and Future Work}

The proposed bilevel-optimized INR still requires coil sensitivity maps  
estimated from a low-resolution GRE pre-scan or the center of k-space. Although 
the method is faster than most self-supervised reconstruction approaches, it remains 
slower than SENSE, GRAPPA, and similar to other model-based compressed sensing algorithms 
if they are also implemented in PyTorch with GPU support and INR is trained from scratch. 
The time difference is only a few seconds for 2D reconstructions but can become 
significant for higher-dimensional datasets. 
One potential solution is to use a model-based reconstruction as a warm start for the INR.

Another challenge arises from the high dimensionality of 3D or dynamic imaging,
which can require excessive memory use during training. 
Unlike prior INR applications (e.g., gigapixel image fitting \cite{instantngp}), 
where the loss is computed in the image domain and coordinates can be 
divided into smaller batches,
MRI has a forward model involving a FFT
that is most efficient
when the entire image is used at each iteration.
Consequently, the batch size should be 1,
and all coordinates should be fed into the MLP at once. 
A region-of-interest (ROI) mask can reduce the coordinate space,
but may still be inadequate for large 3D volumes.

Finally, although bilevel optimization with weight decay enhances noise robustness,
the powerful function-fitting capacity of INR can still yield noisy reconstructions
when the undersampled k-space data have low SNR or a high acceleration factor.
Improving the noise robustness of INR-based MRI reconstruction
without additional regularizers or denoisers
remains an open research question, especially critical for applications 
like diffusion MRI where SNR inherently limits the reconstruction quality.

\section*{Acknowledgment}
This study was supported by NIH grants R37CA263583, R01CA284172, and Siemens Healthineers.

\bibliographystyle{IEEEtran}
\bibliography{mybib}

\end{document}